\documentclass[twocolumn]{aastex63}

\usepackage{amsmath}
\usepackage{amssymb}
\usepackage{amsthm}
\usepackage{graphicx}
\usepackage[colorinlistoftodos]{todonotes}
\usepackage{natbib}

\usepackage{lineno}
\usepackage{soul}
\usepackage{subfigure}
\usepackage{makecell}
\usepackage{threeparttable}
\definecolor{darkgreen}{rgb}{0.05,0.3,0.05}

\begin{document}

\title{A DESGW Search for the Electromagnetic Counterpart to the LIGO/Virgo Gravitational Wave Binary Neutron Star Merger Candidate S190510g}




\author[0000-0001-9578-6322]{A.~Garcia}
\affiliation{Brandeis University, Physics Department, 415 South Street, Waltham MA 02453}
\author[0000-0002-7016-5471]{R.~Morgan}
\affiliation{Physics Department, 2320 Chamberlin Hall, University of Wisconsin-Madison, 1150 University Avenue Madison, WI  53706-1390}
\author[0000-0001-6718-2978]{K.~Herner}
\affiliation{Fermi National Accelerator Laboratory, P. O. Box 500, Batavia, IL 60510, USA}
\author[0000-0002-6011-0530]{A.~Palmese}
\affiliation{Fermi National Accelerator Laboratory, P. O. Box 500, Batavia, IL 60510, USA}
\affiliation{Kavli Institute for Cosmological Physics, University of Chicago, Chicago, IL 60637, USA}
\author[0000-0001-6082-8529]{M.~Soares-Santos}
\affiliation{Brandeis University, Physics Department, 415 South Street, Waltham MA 02453}
\author[0000-0002-0609-3987]{J.~Annis}
\affiliation{Fermi National Accelerator Laboratory, P. O. Box 500, Batavia, IL 60510, USA}
\author{D.~Brout}
\affiliation{Department of Physics and Astronomy, University of Pennsylvania, Philadelphia, PA 19104, USA}
\author[0000-0003-4341-6172]{A.~K.~Vivas}
\affiliation{Cerro Tololo  Inter-American Observatory, NSF’s National Optical-Infrared Astronomy Research Laboratory, Casilla 603, La Serena,
Chile}
\author[0000-0001-8251-933X]{A.~Drlica-Wagner}
\affiliation{Department of Astronomy and Astrophysics, University of Chicago, Chicago, IL 60637, USA}
\affiliation{Fermi National Accelerator Laboratory, P. O. Box 500, Batavia, IL 60510, USA}
\affiliation{Kavli Institute for Cosmological Physics, University of Chicago, Chicago, IL 60637, USA}
\author[0000-0003-3402-6164]{L. ~Santana-Silva}
\affiliation{Federal University of Rio de Janeiro,Valongo Observatory}
\author[0000-0001-7211-5729]{D.~L.~Tucker}
\affiliation{Fermi National Accelerator Laboratory, P. O. Box 500, Batavia, IL 60510, USA}
\author[0000-0002-7069-7857]{S.~Allam}
\affiliation{Fermi National Accelerator Laboratory, P. O. Box 500, Batavia, IL 60510, USA}
\author[0000-0001-8653-7738]{M. ~Wiesner}
\affiliation{Benedictine University}
\author[0000-0002-9370-8360]{J.~Garc\'ia-Bellido}
\affiliation{Instituto de Fisica Teorica UAM/CSIC, Universidad Autonoma de Madrid, 28049 Madrid, Spain}
\author[0000-0003-2524-5154]{M.~S.~S.~Gill}
\affiliation{SLAC National Accelerator Laboratory, Menlo Park, CA 94025, USA}
\author{M.~Sako}
\affiliation{Department of Physics and Astronomy, University of Pennsylvania, Philadelphia, PA 19104, USA}
\author[0000-0003-3221-0419]{R.~Kessler}
\affiliation{Department of Astronomy and Astrophysics, University of Chicago, Chicago, IL 60637, USA}
\affiliation{Kavli Institute for Cosmological Physics, University of Chicago, Chicago, IL 60637, USA}
\author[0000-0002-4213-8783]{T.~M.~Davis}
\affiliation{School of Mathematics and Physics, University of Queensland,  Brisbane, QLD 4072, Australia}
\author{D.~Scolnic}
\affiliation{Department of Physics, Duke University Durham, NC 27708, USA}

\author{J. ~Casares}
\affiliation{Instituto de Astrofisica de Caanarias (IAC)}
\author{H.~Chen}
\affiliation{Kavli Institute for Cosmological Physics, University of Chicago, Chicago, IL 60637, USA}
\author[0000-0003-1949-7638]{C.~Conselice}
\affiliation{University of Nottingham, School of Physics and Astronomy, Nottingham NG7 2RD, UK}
\author[0000-0001-5703-2108]{J. ~Cooke}
\affiliation{Swinburne University}
\author{Z.~Doctor}
\affiliation{Kavli Institute for Cosmological Physics, University of Chicago, Chicago, IL 60637, USA}
\author{R.~J.~Foley}
\affiliation{Santa Cruz Institute for Particle Physics, Santa Cruz, CA 95064, USA}
\author{J. ~Horvath}
\affiliation{IAG-USP/Brazil}
\author{D. ~A. ~Howell}
\affiliation{Las Cumbres Observatory}
\author{C. ~D. ~Kilpatrick}
\affiliation{UCSC}
\author[0000-0003-1731-0497]{C.~Lidman}
\affiliation{The Research School of Astronomy and Astrophysics, Australian National University, ACT 2601, Australia}
\author[0000-0002-5115-6377]{F.~Olivares~E.}
\affiliation{Instituto de Astronom\'{\i}a y Ciencias Planetarias, Universidad de Atacama, Copayapu 485, Copiap\'o, Chile}
\author{F.~Paz-Chinch\'{o}n}
\affiliation{Institute of Astronomy, University of Cambridge, Madingley Road, Cambridge CB3 0HA, UK}
\affiliation{National Center for Supercomputing Applications, 1205 West Clark St., Urbana, IL 61801, USA}
\author{J.~Pineda-G.}
\affiliation{Departamento de Ciencias F\'isicas, Universidad Andr\'es Bello, Avda. República 252, Santiago de Chile}
\author{J.~Quirola-V\'asquez}
\affiliation{Instituto de Astrof\'isica, Pontificia Universidad Cat\'olica de Chile, Casilla 306, Santiago 22, ChileMillennium}
\affiliation{Institute of Astrophysics (MAS), Nuncio Monse$\tilde{n}$or S\'otero Sanz 100, Providencia, Santiago, Chile}
\author[0000-0002-4410-5387]{A. ~Rest}
\affiliation{STScI/JHU}
\author[0000-0001-5399-0114]{N.~Sherman}
\affiliation{Brandeis University, Physics Department, 415 South Street, Waltham MA 02453}

\author{T.~M.~C.~Abbott}
\affiliation{Cerro Tololo  Inter-American Observatory, NSF’s National Optical-Infrared Astronomy Research Laboratory, Casilla 603, La Serena,
Chile}
\author{M.~Aguena}
\affiliation{Departamento de F\'isica Matem\'atica, Instituto de F\'isica, Universidade de S\~ao Paulo, CP 66318, S\~ao Paulo, SP, 05314-970, Brazil}
\affiliation{Laborat\'orio Interinstitucional de e-Astronomia - LIneA, Rua Gal. Jos\'e Cristino 77, Rio de Janeiro, RJ - 20921-400, Brazil}
\author{S.~Avila}
\affiliation{Instituto de Fisica Teorica UAM/CSIC, Universidad Autonoma de Madrid, 28049 Madrid, Spain}
\author{E.~Bertin}
\affiliation{CNRS, UMR 7095, Institut d'Astrophysique de Paris, F-75014, Paris, France}
\affiliation{Sorbonne Universit\'es, UPMC Univ Paris 06, UMR 7095, Institut d'Astrophysique de Paris, F-75014, Paris, France}
\author{S.~Bhargava}
\affiliation{Department of Physics and Astronomy, Pevensey Building, University of Sussex, Brighton, BN1 9QH, UK}
\author[0000-0002-8458-5047]{D.~Brooks}
\affiliation{Department of Physics \& Astronomy, University College London, Gower Street, London, WC1E 6BT, UK}
\author{D.~L.~Burke}
\affiliation{Kavli Institute for Particle Astrophysics \& Cosmology, P. O. Box 2450, Stanford University, Stanford, CA 94305, USA}
\affiliation{SLAC National Accelerator Laboratory, Menlo Park, CA 94025, USA}
\author[0000-0003-3044-5150]{A.~Carnero~Rosell}
\affiliation{}
\author[0000-0002-4802-3194]{M.~Carrasco~Kind}
\affiliation{Department of Astronomy, University of Illinois at Urbana-Champaign, 1002 W. Green Street, Urbana, IL 61801, USA}
\affiliation{National Center for Supercomputing Applications, 1205 West Clark St., Urbana, IL 61801, USA}
\author[0000-0002-3130-0204]{J.~Carretero}
\affiliation{Institut de F\'{\i}sica d'Altes Energies (IFAE), The Barcelona Institute of Science and Technology, Campus UAB, 08193 Bellaterra (Barcelona) Spain}
\author{M.~Costanzi}
\affiliation{INAF-Osservatorio Astronomico di Trieste, via G. B. Tiepolo 11, I-34143 Trieste, Italy}
\affiliation{Institute for Fundamental Physics of the Universe, Via Beirut 2, 34014 Trieste, Italy}
\author{L.~N.~da Costa}
\affiliation{Laborat\'orio Interinstitucional de e-Astronomia - LIneA, Rua Gal. Jos\'e Cristino 77, Rio de Janeiro, RJ - 20921-400, Brazil}
\affiliation{Observat\'orio Nacional, Rua Gal. Jos\'e Cristino 77, Rio de Janeiro, RJ - 20921-400, Brazil}
\author[0000-0002-0466-3288]{S.~Desai}
\affiliation{Department of Physics, IIT Hyderabad, Kandi, Telangana 502285, India}
\author[0000-0002-8357-7467]{H.~T.~Diehl}
\affiliation{Fermi National Accelerator Laboratory, P. O. Box 500, Batavia, IL 60510, USA}
\author[0000-0002-8134-9591]{J.~P.~Dietrich}
\affiliation{Faculty of Physics, Ludwig-Maximilians-Universit\"at, Scheinerstr. 1, 81679 Munich, Germany}
\author{P.~Doel}
\affiliation{Department of Physics \& Astronomy, University College London, Gower Street, London, WC1E 6BT, UK}
\author{S.~Everett}
\affiliation{Santa Cruz Institute for Particle Physics, Santa Cruz, CA 95064, USA}
\author[0000-0002-2367-5049]{B.~Flaugher}
\affiliation{Fermi National Accelerator Laboratory, P. O. Box 500, Batavia, IL 60510, USA}
\author{P.~Fosalba}
\affiliation{Institut d'Estudis Espacials de Catalunya (IEEC), 08034 Barcelona, Spain}
\affiliation{Institute of Space Sciences (ICE, CSIC),  Campus UAB, Carrer de Can Magrans, s/n,  08193 Barcelona, Spain}
\author{D.~Friedel}
\affiliation{National Center for Supercomputing Applications, 1205 West Clark St., Urbana, IL 61801, USA}
\author[0000-0003-4079-3263]{J.~Frieman}
\affiliation{Fermi National Accelerator Laboratory, P. O. Box 500, Batavia, IL 60510, USA}
\affiliation{Kavli Institute for Cosmological Physics, University of Chicago, Chicago, IL 60637, USA}
\author[0000-0001-9632-0815]{E.~Gaztanaga}
\affiliation{Institut d'Estudis Espacials de Catalunya (IEEC), 08034 Barcelona, Spain}
\affiliation{Institute of Space Sciences (ICE, CSIC),  Campus UAB, Carrer de Can Magrans, s/n,  08193 Barcelona, Spain}
\author[0000-0001-6942-2736]{D.~W.~Gerdes}
\affiliation{Department of Astronomy, University of Michigan, Ann Arbor, MI 48109, USA}
\affiliation{Department of Physics, University of Michigan, Ann Arbor, MI 48109, USA}
\author[0000-0003-3270-7644]{D.~Gruen}
\affiliation{Department of Physics, Stanford University, 382 Via Pueblo Mall, Stanford, CA 94305, USA}
\affiliation{Kavli Institute for Particle Astrophysics \& Cosmology, P. O. Box 2450, Stanford University, Stanford, CA 94305, USA}
\affiliation{SLAC National Accelerator Laboratory, Menlo Park, CA 94025, USA}
\author{R.~A.~Gruendl}
\affiliation{Department of Astronomy, University of Illinois at Urbana-Champaign, 1002 W. Green Street, Urbana, IL 61801, USA}
\affiliation{National Center for Supercomputing Applications, 1205 West Clark St., Urbana, IL 61801, USA}
\author[0000-0003-3023-8362]{J.~Gschwend}
\affiliation{Laborat\'orio Interinstitucional de e-Astronomia - LIneA, Rua Gal. Jos\'e Cristino 77, Rio de Janeiro, RJ - 20921-400, Brazil}
\affiliation{Observat\'orio Nacional, Rua Gal. Jos\'e Cristino 77, Rio de Janeiro, RJ - 20921-400, Brazil}
\author[0000-0003-0825-0517]{G.~Gutierrez}
\affiliation{Fermi National Accelerator Laboratory, P. O. Box 500, Batavia, IL 60510, USA}
\author{S.~R.~Hinton}
\affiliation{School of Mathematics and Physics, University of Queensland,  Brisbane, QLD 4072, Australia}
\author{D.~L.~Hollowood}
\affiliation{Santa Cruz Institute for Particle Physics, Santa Cruz, CA 95064, USA}
\author[0000-0002-6550-2023]{K.~Honscheid}
\affiliation{Center for Cosmology and Astro-Particle Physics, The Ohio State University, Columbus, OH 43210, USA}
\affiliation{Department of Physics, The Ohio State University, Columbus, OH 43210, USA}
\author[0000-0001-5160-4486]{D.~J.~James}
\affiliation{Center for Astrophysics $\vert$ Harvard \& Smithsonian, 60 Garden Street, Cambridge, MA 02138, USA}
\author[0000-0003-0120-0808]{K.~Kuehn}
\affiliation{Australian Astronomical Optics, Macquarie University, North Ryde, NSW 2113, Australia}
\affiliation{Lowell Observatory, 1400 Mars Hill Rd, Flagstaff, AZ 86001, USA}
\author[0000-0003-2511-0946]{N.~Kuropatkin}
\affiliation{Fermi National Accelerator Laboratory, P. O. Box 500, Batavia, IL 60510, USA}
\author[0000-0002-1134-9035]{O.~Lahav}
\affiliation{Department of Physics \& Astronomy, University College London, Gower Street, London, WC1E 6BT, UK}
\author{M.~Lima}
\affiliation{Departamento de F\'isica Matem\'atica, Instituto de F\'isica, Universidade de S\~ao Paulo, CP 66318, S\~ao Paulo, SP, 05314-970, Brazil}
\affiliation{Laborat\'orio Interinstitucional de e-Astronomia - LIneA, Rua Gal. Jos\'e Cristino 77, Rio de Janeiro, RJ - 20921-400, Brazil}
\author[0000-0001-9856-9307]{M.~A.~G.~Maia}
\affiliation{Laborat\'orio Interinstitucional de e-Astronomia - LIneA, Rua Gal. Jos\'e Cristino 77, Rio de Janeiro, RJ - 20921-400, Brazil}
\affiliation{Observat\'orio Nacional, Rua Gal. Jos\'e Cristino 77, Rio de Janeiro, RJ - 20921-400, Brazil}
\author{M.~March}
\affiliation{Department of Physics and Astronomy, University of Pennsylvania, Philadelphia, PA 19104, USA}
\author[0000-0003-0710-9474]{J.~L.~Marshall}
\affiliation{George P. and Cynthia Woods Mitchell Institute for Fundamental Physics and Astronomy, and Department of Physics and Astronomy, Texas A\&M University, College Station, TX 77843,  USA}
\author[0000-0002-1372-2534]{F.~Menanteau}
\affiliation{Department of Astronomy, University of Illinois at Urbana-Champaign, 1002 W. Green Street, Urbana, IL 61801, USA}
\affiliation{National Center for Supercomputing Applications, 1205 West Clark St., Urbana, IL 61801, USA}
\author[0000-0002-6610-4836]{R.~Miquel}
\affiliation{Instituci\'o Catalana de Recerca i Estudis Avan\c{c}ats, E-08010 Barcelona, Spain}
\affiliation{Institut de F\'{\i}sica d'Altes Energies (IFAE), The Barcelona Institute of Science and Technology, Campus UAB, 08193 Bellaterra (Barcelona) Spain}
\author[0000-0003-2120-1154]{R.~L.~C.~Ogando}
\affiliation{Laborat\'orio Interinstitucional de e-Astronomia - LIneA, Rua Gal. Jos\'e Cristino 77, Rio de Janeiro, RJ - 20921-400, Brazil}
\affiliation{Observat\'orio Nacional, Rua Gal. Jos\'e Cristino 77, Rio de Janeiro, RJ - 20921-400, Brazil}
\author[0000-0002-2598-0514]{A.~A.~Plazas}
\affiliation{Department of Astrophysical Sciences, Princeton University, Peyton Hall, Princeton, NJ 08544, USA}
\author[0000-0002-9328-879X]{A.~K.~Romer}
\affiliation{Department of Physics and Astronomy, Pevensey Building, University of Sussex, Brighton, BN1 9QH, UK}
\author[0000-0001-5326-3486]{A.~Roodman}
\affiliation{Kavli Institute for Particle Astrophysics \& Cosmology, P. O. Box 2450, Stanford University, Stanford, CA 94305, USA}
\affiliation{SLAC National Accelerator Laboratory, Menlo Park, CA 94025, USA}
\author[0000-0002-9646-8198]{E.~Sanchez}
\affiliation{Centro de Investigaciones Energ\'eticas, Medioambientales y Tecnol\'ogicas (CIEMAT), Madrid, Spain}
\author{V.~Scarpine}
\affiliation{Fermi National Accelerator Laboratory, P. O. Box 500, Batavia, IL 60510, USA}
\author[0000-0001-9504-2059]{M.~Schubnell}
\affiliation{Department of Physics, University of Michigan, Ann Arbor, MI 48109, USA}
\author{S.~Serrano}
\affiliation{Institut d'Estudis Espacials de Catalunya (IEEC), 08034 Barcelona, Spain}
\affiliation{Institute of Space Sciences (ICE, CSIC),  Campus UAB, Carrer de Can Magrans, s/n,  08193 Barcelona, Spain}
\author[0000-0002-1831-1953]{I.~Sevilla-Noarbe}
\affiliation{Centro de Investigaciones Energ\'eticas, Medioambientales y Tecnol\'ogicas (CIEMAT), Madrid, Spain}
\author[0000-0002-3321-1432]{M.~Smith}
\affiliation{School of Physics and Astronomy, University of Southampton,  Southampton, SO17 1BJ, UK}
\author[0000-0002-7047-9358]{E.~Suchyta}
\affiliation{Computer Science and Mathematics Division, Oak Ridge National Laboratory, Oak Ridge, TN 37831}
\author{M.~E.~C.~Swanson}
\affiliation{National Center for Supercomputing Applications, 1205 West Clark St., Urbana, IL 61801, USA}
\author[0000-0003-1704-0781]{G.~Tarle}
\affiliation{Department of Physics, University of Michigan, Ann Arbor, MI 48109, USA}
\author{D.~Thomas}
\affiliation{Institute of Cosmology and Gravitation, University of Portsmouth, Portsmouth, PO1 3FX, UK}
\author{T.~N.~Varga}
\affiliation{Max Planck Institute for Extraterrestrial Physics, Giessenbachstrasse, 85748 Garching, Germany}
\affiliation{Universit\"ats-Sternwarte, Fakult\"at f\"ur Physik, Ludwig-Maximilians Universit\"at M\"unchen, Scheinerstr. 1, 81679 M\"unchen, Germany}
\author[0000-0002-7123-8943]{A.~R.~Walker}
\affiliation{Cerro Tololo  Inter-American Observatory, NSF’s National Optical-Infrared Astronomy Research Laboratory, Casilla 603, La Serena,
Chile}
\author[0000-0002-8282-2010]{J.~Weller}
\affiliation{Max Planck Institute for Extraterrestrial Physics, Giessenbachstrasse, 85748 Garching, Germany}
\affiliation{Universit\"ats-Sternwarte, Fakult\"at f\"ur Physik, Ludwig-Maximilians Universit\"at M\"unchen, Scheinerstr. 1, 81679 M\"unchen, Germany}

\collaboration{1000}{(DES Collaboration)}


\email{alyssag94@brandeis.edu}

\begin{abstract}

We present the results from a search for the electromagnetic counterpart of the LIGO/Virgo event S190510g using the Dark Energy Camera (DECam). S190510g is a binary neutron star (BNS) merger candidate of moderate significance detected at a distance of 227$\pm$92 Mpc and localized within an area of 31 (1166) square degrees at 50\% (90\%) confidence. While this event was later classified as likely non-astrophysical in nature within 30 hours of the event, our short latency search and discovery pipeline identified 11 counterpart candidates, all of which appear consistent with supernovae following offline analysis and spectroscopy by other instruments. Later reprocessing of the images enabled the recovery of 6 more candidates. Additionally, we implement our candidate selection procedure on simulated kilonovae and supernovae under DECam observing conditions (e.g., seeing, exposure time) with the intent of quantifying our search efficiency and making informed decisions on observing strategy for future similar events. This is the first BNS counterpart search to employ a comprehensive simulation-based efficiency study. We find that using the current follow-up strategy, there would need to be 19 events similar to S190510g for us to have a 99\% chance of detecting an optical counterpart, assuming a GW170817-like kilonova. We further conclude that optimization of observing plans, which should include preference for deeper images over multiple color information, could result in up to a factor of 1.5 reduction in the total number of followups needed for discovery.

\end{abstract}



\reportnum{DES-2019-0478}
\reportnum{FERMILAB-PUB-20-255-AE}


\section{Introduction}

Binary neutron star mergers such as GW170817 \citep{ligobns}, in which both a gravitational wave and its electromagnetic counterpart were detected, can be used for measurements such as an independent calculation of the Hubble constant (\citealt{schutz,delpozzo,2017Natur.551...85A}; \citealt*{darksiren1}), or even to probe the growth of structure from peculiar velocities \citep{palmese20}. For this reason, the Dark Energy Survey (DES; \citealt{2016MNRAS.460.1270D}) launched the gravitational wave (GW) program (DESGW) in 2015. This program works to quickly identify the optical counterparts to GW events, particularly the kilonovae (KN) expected from binary neutron star mergers 
to be used for cosmology. These transients are produced by the radioactive decay of r-process nuclei synthesized in the merger ejecta and are predicted to be rapidly fading (typically within a few days; \citealt{kasen2017origin}), thus require follow up shortly after the announcement of the trigger. The identification of the KN associated with GW170817 \citep{soares2017electromagnetic,abbott2017multi, coulter2017swope, cowperthwaite2017electromagnetic, evans2017swift, andreoni2017follow, hu2017optical, utsumi2017j, valenti2017discovery, shappee2017early, mccully2017rapid, kasliwal2017illuminating} in the nearby galaxy NGC 4993 \citep{2017ApJ...849L..34P, gw170817_7} is an example of DESGW's  ability to quickly identify these transients and characterize them. 

The Laser Interferometer Gravitational-wave Observatory (LIGO; \citealt{advanced_ligo}) and Virgo \citep{virgo_interferometer} Collaboration (LVC) recently completed its third observing run (O3), from April 2019 through March 2020. During this time, there were 56 publicly reported GW candidates, 14 of which thought to have been originated from binary systems where at least one object's mass was consistent with a neutron star.
In previous LVC observing runs, 11 total events (confirmed GW and marginal triggers) were observed within roughly 14 months \citep{abbott2019gwtc}. The increase in number of GW triggers in O3 relative to earlier runs is due to a significant increase in sensitivity \citep{abbott2018prospects} for all three detectors. This also means that, while the LVC network are detecting more events, many of these events are further away than the first two runs and poorly localized. 

While the optical counterparts of these events are challenging to detect with small telescopes, the Dark Energy Camera (DECam; \citealt{flaugher2015dark}) is optimally suited to find these sources (as shown in \citealt{soares2016dark,soares2017electromagnetic}). DECam's 4m primary mirror allows us to quickly cover large areas of sky down to the limits required to detect EM counterparts of LVC's sources. DECam has been widely used by the community to search for GW counterpart searches. For example, the Global Relay of Observatories Watching Transients Happen (GROWTH; \citealt{goldstein2019growth_190426,andreoni2019growth190814}) collaboration has employed DECam data wide-area searches for several GW candidate events, including S190510g \citep{andreoni2019growth}. None of the search teams have identified a new GW event counterpart since GW170817. In order to interpret the lack of detection, and make informed decisions for future searches, an in-depth analysis  including  simulation-based efficiency study is required. While general studies using average depth have been published \citep{carracedo2020detectability}, this is the first study to utilize simulations that include the impact of observing conditions and observation plan. Such an analysis had not been published until now. See also our companion paper on S190814bv, a neutron star black hole candidate event (\citealt{Morgan_0814bv}), and a standard siren analysis using S190814bv with DES galaxies (Palmese et al., in prep.). 

In this paper we present the DESGW search for the KN counterpart to LVC candidate event S190510g. 
We include results from simulations that allow us to make quantitative statements about sensitivity in light of realistic observing conditions and strategy choices.
The paper is organized as follows: in Section 2 we summarize the search and discovery pipeline used by the DESGW program and give an overview of the candidates discovered; in Section 3 we discuss the method for detecting candidates and for using simulated supernova (SN) and KN light curves; in Section 4 we present the results of our search and discovery pipeline as well as simulation analysis; Section 5 discusses our search efficiency and implications for future follow up strategies; finally, we summarize our analysis in Section 6.

\section{Data}

\subsection{The LIGO/Virgo event S190510g}

All three LVC detectors (LIGO Livingston, LIGO Hanford, and Virgo) recorded the event, with a 98\% initial probability of being a binary neutron star (BNS) event, a 2\% probability of having a non-astrophysical origin, and a false alarm rate of 1 per 37 years.  The 50\% (90\%) confidence regions spanned 575 deg$^2$ (3462 deg$^2$)  in the initial LVC bayestar localization map. At 10:08:19 UTC on May 10, the LVC released an updated map from the LaLInference pipeline \citep{LaLInference2015}, decreasing the $50\%$ and $90\%$ confidence regions to 31 deg$^2$ and 1166 deg$^2$ respectively, and refined the distance estimate to  $227 \pm 92$ Mpc, or $z=0.05 \pm 0.02$ (using flat $\Lambda$CDM cosmology with $H_0 = 70$ km/s/Mpc and $\Omega_m = 0.3$) \citep{2019GCN.24489....1L}. On May 10, 20:43:51 UTC the classification of the nature of the event was updated to $85\%$ BNS and $15\%$ non-astrophysical. Finally at 20:18:44 UTC on May 11, the LVC updated this probability to being non-astrophysical origin at 58\% and of a BNS to 42\% as well as updating the false alarm rate to 1 in 3.6 years.

\subsection{DECam Observations}

\begin{figure}[htp]
\centering
\includegraphics[width=1.0\linewidth]{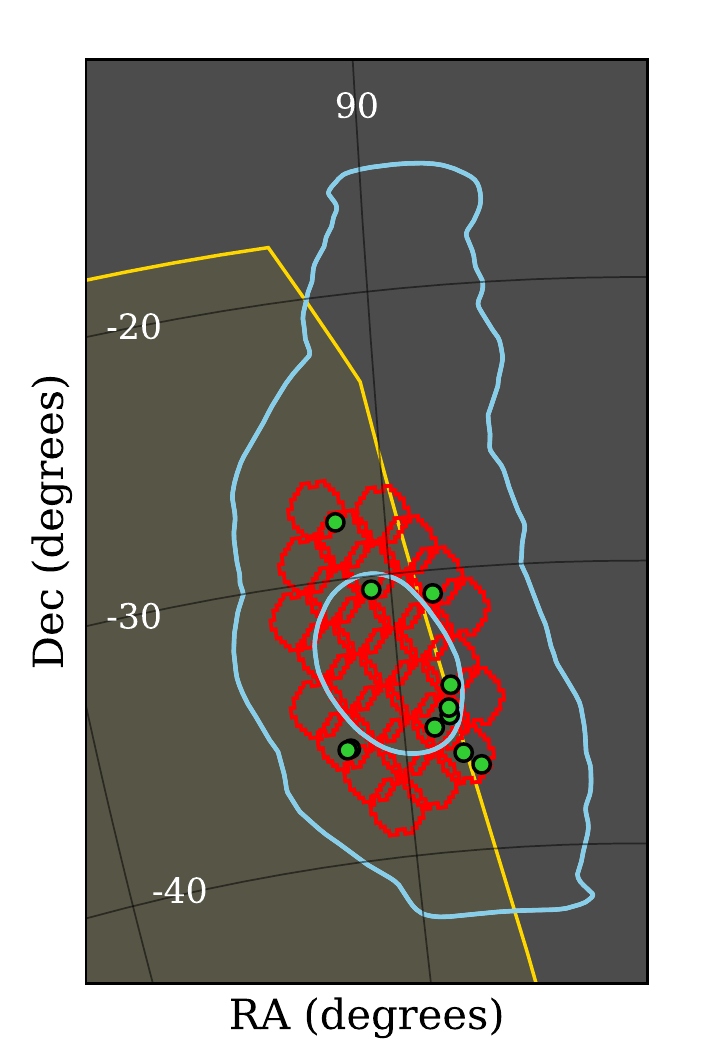}
\caption{Summary of candidates identified by the DESGW short latency pipeline and exposures used for analysis. Here we show one of the three distinct localization regions from the LALInference map. The blue contours show a single region of the LVC $90\%$ and $50\%$ localization contours, the yellow line and shaded area is the DES footprint, and green dots indicate our candidates.
A single pointing of DECam exposures are shown as red hexes which cover $\sim$84 deg$^2$ total and contain $\sim$65\% of the total probability.}\label{fig:hexandcands}
\end{figure}

DECam was used for two nights to conduct target-of-opportunity imaging of the LIGO/Virgo GW compact binary merger candidate S190510g \citep{2019GCN.24442....1L}. Since the initial classification of S190510g was a BNS merger with high probability, the GROWTH collaboration chose to trigger DECam (NOAO proposal 2019A-0205). All exposures from this proposal were immediately made public \citep{andreoni2019growth}. GROWTH initiated EM follow up on May 10th at 06:00:25.488 UTC. The observing plan on this evening was based on the original LVC bayestar probability map. The updated LVC LALInference map disfavored most of the region observed on the first night. As a result GROWTH prepared a new observing plan for the second night \citep{andreoni2019growth}. This plan consisted of observing for $\sim$1.5 hrs beginning at 22:53:04 UTC on May 10. 80 exposures total were taken in the $g$, $r$, and $z$ bands for 40 seconds each. Each filter visited roughly same area of the sky, approximately 30 minutes apart, in order to eliminate moving objects. The $10\sigma$ depths for each band are $m_z = 20.58$ mag, $m_r = 21.72$ mag, and $m_g = 21.67$ mag, where the average seeing was 1.33 arcsec, the average airmass was 1.71, and the average attenuation due to cloud was 4\%. These observations covered $\sim65\%$ of the probability region, as shown in Figure~\ref{fig:hexandcands}. Plans to follow up this event for a third night were retracted due to the updated classification probability of this event. Our analysis uses only the exposures from the second night of observations as to include only the high probability region from the LALInference LVC map.

\section{Methods}
\subsection{Search and Discovery Pipeline}

Images from DECam were downloaded directly to Fermi National Accelerator Laboratory from Cerro Tololo Inter-American Observatory (CTIO) via the National Center for Supercomputing Applications (NCSA). Once the search exposures became available, we immediately started initial image processing, parallelized to run on a CCD by CCD basis, via the DESGW Search and Discovery Pipeline~\citep{details}. The pipeline consists of three major stages: Single-Epoch (SE) processing, difference imaging, and post-processing.
 
SE processing~\citep{2018PASP..130g4501M} consists of image correction and astrometric calibration. This stage uses  {\tt SExtractor} \citep{bertin1996sextractor} to create a list of bright objects in each image, which is fed into {\tt scamp} for astrometric calibration \citep{2006ASPC..351..112B}. We use the GAIA-DR2 catalog \citep{2016A&A...595A...1G,2018A&A...616A...1G} 
in this stage, which allows reductions of DES astrometric uncertainties to below $0.03^{\prime\prime}$. The outputs of SE processing are the inputs to the difference imaging ({\tt diffimg}) stage, designed for the DES supernova {\tt diffimg} pipeline~\citep{kessler2015}, but modified to work with wide-survey images. {\tt Diffimg} subtracts one or more template images from the search image. Template images are taken in the same area of sky before the event, or after the event is expected to have faded if pre-existing templates are unavailable. Our group is able to use all DES images as templates, including those not yet publicly released, as well as all public non-DES DECam images. Combining public DECam data and not-yet-released DES data, we are able to improve on the depth of template images by roughly $\sim\sqrt{2}$ within the DES footprint.
After the  {\tt diffimg} subtraction is complete, post-processing takes candidate objects identified and applies a cut based on the machine learning algorithm ({\tt autoscan}) score ~\citep{goldstein2015,goldstein2015err}. Furthermore, forced photometry (described in \citet{kessler2015}) is applied in this step, as well as host galaxy matching and additional requirements such as detections in multiple bands and/or on multiple nights.
For S190510g, these requirements were only a detection in two or more exposures and a machine learning score adequate to reject non-astrophysical sources. We eliminated known asteroids using the minor planet center, but since a second night of exposures were not taken in the region of interest, we did not eliminate other asteroids from our candidate list.
Finally, the resulting candidate list was vetted via human inspection, as described in Section 3.2. The stamps for each of the DESGW candidates are shown in Figure~\ref{fig:stamps}.

\subsubsection{Pipeline Performance}
Roughly $26\%$ of the image processing jobs took between 0 -- 30$^{\text{min}}$ to complete, $23\%$ took $0.5 - 1^{\text{hr}}$, $22\%$ took $1 - 1.5^{\text{hr}}$, $14\%$ took $1.5 - 2^{\text{hr}}$, while the rest took $> 2^{\text{hr}}$ to complete. The image processing section of our pipeline runs on a parallelized CCD per CCD basis. This means that for the 80 exposures used for this analysis, there were $\sim$5000 jobs total. Post processing also runs  on a CCD per CCD basis after image processing has finished. This step takes $\sim$20 min to finish when running with all exposures. 
We note that this turnaround time is significantly longer than the GROWTH team reported in \citep{andreoni2019growth}. This is likely due to a combination of having, on average, more template images and applying  a more complete correction set in the SE stage, such as correcting for the brighter-fatter effect \citep{bernstein2017instrumental}.

\subsection{Candidate identification}

In total, there were 1165 candidates identified after post-processing. The final candidate list was published in GCN 24480 at 12:24 pm May 11 UTC \citep{2019GCN.24480....1S}. 
The primary cuts for our candidates require no {\tt SExtractor} errors in image processing, such as masking of objects overlapping the transient or inability to measure the flux, and an {\tt autoscan} score of at least 0.9 out of 1.0. This cut found 96 candidates (20 with {\tt autoscan} score $> 0.95$), while the final 11 were selected via visual inspection. The key properties we looked for when performing visual inspection is a host galaxy in the template image, a non-noisy template image, and no regions of over or under-subtraction. We also took into consideration the possibility that the candidate could be an AGN since we are unable to resolve objects that are close to the center of the host galaxy and therefore disfavored stamps where the candidate is not distinguishable from the host galaxy. Further, we note that no candidate from our pipeline is fully dismissed until there is secondary follow-up or enough evidence to definitively categorize the object. For a single night of observations, our goal is to rapidly identify objects that are the most obvious candidates, then refine our search criteria as we observe more epochs. 

Additionally, we matched candidates to hosts and used DES data to measure properties of the host, such as photometric redshift, absolute magnitude, stellar mass, and  star formation rate, as well as the separation of the candidate and the host at the redshift of the nearest potential host galaxy.
Photometric redshifts have been computed using Directional Neighborhood Fitting (DNF; \citealt{dnf}), while the galaxy properties have been computed using the Bayesian Model Averaging method as described in \citet{2019arXiv190308813P}.
The coordinates and other information about each of our candidates can be found in Table 1, and information about their host galaxies is listed in Table 2.

\section{Results}

 \subsection{Candidate Classification}
The first stage of analysis, performed as exposures became available, presented 11 candidates (of which 6 were also detected by GROWTH) that were produced via the DESGW Search and Discovery Pipeline discussed in Section 3. Follow up from other observatories is crucial for determining if a candidate is the GW counterpart through rejection of false positives.
The Korea Microlensing Telescope Network (KMTNet) followed up five of our candidates, desgw-190510a, desgw-190510c, desgw-190510i, desgw-190510j, and desgw-190510k (GCN 24493 and 24529; \citealt{2019GCN.24493....1I,2019GCN.24529....1I}), 
at the KMTNet South Africa (SAAO), Chile (CTIO), and Australia (SSO) stations showing that each of these candidates did not have significant fading over $\sim$1 day, but did show very slow or no fading, therefore deeming these candidates likely supernovae.
 Additionally desgw-190510c was observed by Swift-XRT (GCN 24541; \citealt{2019GCN.24541....1E}), showing no XRT source found, as well as with Magellan (GCN 24511; \citealt{2019GCN.24511....1G}), which found a broad feature consistent with H-$\alpha$ at a redshift of 0.06 and suggests a good match to a Type II SN approximately one week after peak brightness. Finally, desgw-190510h was initially detected by ATLAS on March 13, 2019 and later classified as a Type Ia SN at redshift 0.07 roughly a few days after maximum light by the Spectral Classification of Astronomical Transients (SCAT) survey and desgw190510-b was recorded by Gaia on Jan 30, 2019 and reported as a ``blue hostless transient". This transient can also be seen in previous DES images dating about 2.5 years ago, though with not enough information to classify it with certainty, thus we provide no host information in Table~\ref{tab:gals}. This leaves only 4 candidates, desgw-190510d, e,f, and g, that were not classified by secondary follow-up, and thus still potential counterpart candidates. 
 
 The remaining information about each candidate that can be used to determine if a candidate is viable can be found in Table 2. The table reports photometric redshift, star formation rate, stellar mass, and absolute magnitude of the hosts, computed using DES Year 3 data \citep{dr1}. Furthermore, galaxies are ranked based on their probability of association, which can be computed using the skymap information, the galaxies' position and redshift \citep{2016ApJS..226...10S}, assuming a flat $\Lambda$CDM cosmology with $H_0=70$ km~s$^{-1}$ Mpc$^{-1}$ and $\Omega_m=0.3$.

\begin{figure*}[ht]
\centering
\includegraphics[width=1.0\linewidth]{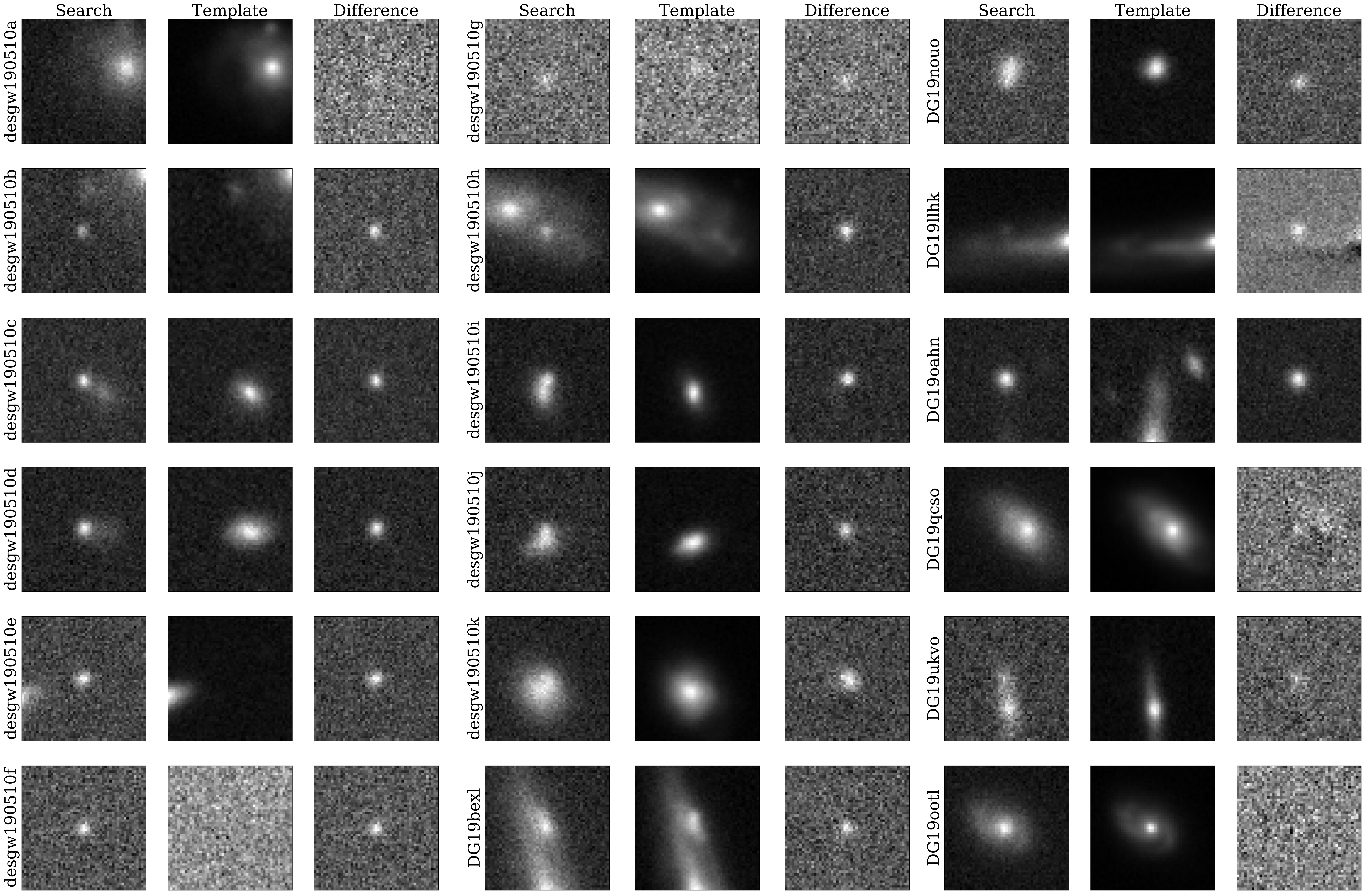}
\caption{Stamps (search, template, and difference images) for all S190510g candidates found by DESGW, including 11 candidates detected in our short latency search (as reported in GCN 24480) and 7 candidates which were first reported by the GROWTH collaboration. Six of the GROWTH candidates were later found by our pipeline, while DG19ootl was not. }\label{fig:stamps}
\end{figure*}

\begin{center}
\begin{table*}[ht] 
\begin{threeparttable}
\caption{Candidates that pass the DESGW pipeline cuts as well as visual inspection from the first stage of analysis. If the candidate matches a candidate from the GROWTH team's candidate list, the GROWTH name is stated. If additional follow up conducted by other telescopes verified the candidate as SN the classification is listed. Those labeled only ``SN" did not have sufficient information for specific classification.
\label{tab:cuts}}
\hspace*{-3cm}
\begin{tabular*}{1.0\linewidth}{@{\extracolsep{\fill}}lccccccc}
\toprule
DES (GROWTH) Name & mag $g$ & mag $r$ & mag $z$ & RA (deg) & DEC (deg) & {\tt autoscan} & Class- \\
 &  &  &  &  &  & Score & ification \\[1ex]
\hline
desgw-190510a &  
22.53 $\pm$ 0.19 & 
20.93 $\pm$ 0.04 & 
20.77 $\pm$ 0.10 &
91.526744 &  
-35.541616 & 
0.950 & SN \\ [1ex]
desgw-190510b & 
21.19 $\pm{}$ 0.05 & 
21.13 $\pm{}$ 0.05 &   
 &  
93.704382 & 
-36.980727 &  
0.950 & \\ [1ex]
desgw-190510c (DG19fqqk) &  
21.72 $\pm{}$ 0.1 & 
20.37 $\pm{}$ 0.02 & 
20.35 $\pm{}$ 0.06 &   
92.851468 &  
-36.517324 & 
0.970 & 
SN II \\ [1ex]
desgw-190510d (DG19nanl) &    
20.36 $\pm{}$ 0.03 &  
19.92 $\pm{}$ 0.2 & 
20.77 $\pm{}$ 0.11 &   
87.311398 & 
-35.955853 & 
0.970 & \\ [1ex]
desgw-190510e (DG19etsk) &    
20.56 $\pm{}$ 0.03 & 
20.66 $\pm{}$ 0.03 & 
20.82 $\pm{}$ 0.09 &   
89.100926 &  
-30.473987 &  
0.970 & \\ [1ex]
desgw-190510f &    
22.16 $\pm{}$ 0.13 &  
21.30 $\pm{}$ 0.05 &   
& 
92.294458 &  
-34.884684 &  
0.970 &\\ [1ex]
desgw-190510g  &    
22.48 $\pm{}$ 0.17 & 
21.92 $\pm{}$ 0.09 &  
&   
92.468923 &  
-34.08657  &
0.963 &\\ [1ex]
desgw-190510h  &     21.23 $\pm{}$ 0.08 &  20.29 $\pm{}$ 0.03 &  20.56 $\pm{}$ 0.07 &   87.762354 &  -27.956502 &  0.960 & SN \\ [1ex]
desgw-190510i (DG19yhhm) &     &  20.15 $\pm{}$ 0.02 & 20.47 $\pm{}$ 0.08 &   91.936973 &  -30.824747 & 0.915 & SN \\ [1ex]
desgw-190510j (DG19zaxn) &   20.65 $\pm{}$ 0.04 &  20.83 $\pm{}$ 0.04 &  &   92.307977 &  -35.149829 & 0.900 & SN \\ [1ex]
desgw-190510k (DG19lcnl) & 20.15 $\pm{}$ 0.03 &  &  19.53 $\pm{}$ 0.03 &   87.146843 &  -35.994357 &  0.920 & SN \\ [1ex]
\hline
\end{tabular*}
\end{threeparttable}
\end{table*}
\end{center}

\begin{center}
\begin{table*}[ht]
\begin{threeparttable}
\caption{Candidate host galaxies' information. Redshifts listed are mean photometric redshifts with one sigma errors. All photometric data, star formation rates (SFR), stellar mass ($M_\star$), and magnitudes are computed using DES Year 3 data. Additionally, the separation (``Sep'') between candidate and host galaxy is calculated using the redshift of the galaxy. Galaxies are ranked based on their position in the sky and redshift, using the information provided in the skymap. Log indicates a logarithm in base 10. 
\label{tab:gals}}
\hspace*{-3cm}
\begin{tabular*}{1.0\linewidth}{@{\extracolsep{\fill}}lccccccc}
\toprule
Name & Host Gal. Name & Sep & $z$ & Log($M_\star$) & Log(SFR) & $M_i$ & Rank \\ 
 & & [kpc] & & Log([$M_\odot$])& Log([M$_\odot$/yr])& & \\ [1ex]
\hline

desgw-190510a & 2MASS J06060625-3532351 & 32.1 & 0.106 $\pm$ 0.004 & 11.069$^{+0.19}_{-0.05}$ & -0.190 & -23.043 &3 \\ [1ex] 
desgw-190510b &  &       &   &  &  &\\ [1ex] 
desgw-190510c &  DES J061124.4562-363104.494 & 97.7 & 0.202 $\pm$ 0.098 & 9.028$^{+0.055}_{-0.060}$ & -0.462 & -20.071 &9 \\ [1ex]
desgw-190510d & WISEA J054914.81-355724.3 & 5.0 & 0.130 $\pm$ 0.021 & 9.388$^{+0.14}_{-0.09}$ & -0.502 & -19.33 &4\\ [1ex]
desgw-190510e & WISEA J055624.41-302817.8 & 24.6 & 0.163 $\pm$ 0.005 & 10.829$^{+0.05}_{-0.12}$ & -0.390 & -21.796 & 8  \\ [1ex]
desgw-190510f & 2MASS J06091226-3452506 & 68.9 & 0.168 $\pm$ 0.003 & 11.200$^{+0.034}_{-0.034}$ & -0.156 & -22.685 &7 \\ [1ex]
desgw-190510g & DES J060952.4784-340540.704 & 218.3 & 0.575 $\pm$ 0.174 & 9.517$^{+0.17}_{-0.14}$ & 0.259 & -20.557 &10 \\ [1ex]
desgw-190510h & 2MASS J05510277-2757201 & 4.3 & 0.049 $\pm$ 0.002 & 10.004$^{+0.034}_{-0.052}$ & -1.388 & -20.917 & 1\\ [1ex]
desgw-190510i & WISEA J060745.00-304928.7 & 55.7 & 0.193 $\pm$ 0.019 & 10.341$^{+0.052}_{-0.041}$ & 0.387 & -21.539 &6 \\ [1ex]
desgw-190510j & WISEA J060914.02-350858.5 & 3.1 & 0.134 $\pm$ 0.014 & 9.69$^{+0.14}_{-0.13}$ & -0.461 & -19.927 &5\\[1ex]
desgw-190510k & 2MASS J05483537-3559390 & 1.9 & 0.067 $\pm$ 0.002 & 9.829$^{+0.09}_{-0.20}$ & -0.272 & -20.627 & 2 \\ [1ex]
\hline
\end{tabular*}
\end{threeparttable}
\end{table*}
\end{center}
 
 \subsection{Recovered Candidates}
  Using the same exposures, the GROWTH collaboration reported a list of 13 candidates (GCN 24467; \citealt{2019GCN.24467....1A}). Seven of the GROWTH candidates were not listed in the initial DESGW candidate list reported in GCN 24480. Candidates DG19bexl and DG19nouo were found in the final stages by our automated pipeline; DG19bexl did not pass the {\tt autoScan} score cut ($\ge 0.9$) and DG19nouo was rejected due to visual inspection. Candidates DS19qcso and DG19llhk both had a detection in a single exposure, where two were required to be picked up as a candidate. The overlapping search exposures for these candidates failed in the {\tt HOTPANTS} step of our pipeline. Reprocessing of these exposures with an updated (current) version of the DESGW  pipeline did identify these candidates. Similarly, candidates DG19ukvo and DG19oahn were not found in our initial processing of the event due to {\tt HOTPANTS } errors in all exposures. DG19oahn was later found in reprocessing, while DG19ukvo continued to have processing failures in 2 out of the 3 exposures. The fraction of missing candidates is consistent with the overall failure rate of $28\%$ for all jobs that were submitted on that night, where $\sim15\%$ of total jobs failed due to issues in {\tt HOTPANTS}. These failures are largely due to the observing conditions described in Section 2.2.
  Finally, candidate DG19ootl was never found in our pipeline. The templates used for this exposure were taken from not yet publicly available DES images and thus did not show any source in the difference image. Candidates, including those initially detected only by GROWTH, are shown in Figure~\ref{fig:stamps}.

\section{Discussion}

\subsection{Understanding Search Efficiency}

To better understand our search efficiency, we performed an off-line analysis using SuperNova ANAlysis software suite (SNANA)   \citep{kessler2009snana}. These simulations produce SN \& KN lightcurves as they would be observed during our observations. Each KN simulation randomly assigns an ejecta mass, ejecta velocity, and lanthinide fraction based on the Kasen KN model \citep{kasen2017origin}. The SN simulations use the SALT2 model for SN Ia \citep{guy2010salt2} and templates for the core collapse SN (SN CC) are taken from
\citet{Kessler_2010} and \cite{Jones_2018}.

Using these simulations, we computed the detection efficiency for each KN model given our observing conditions, the results of which are shown in Figure~\ref{fig:robplot}. The KN simulations used for this analysis produce events that use a distance distribution consistent with that reported by the LVC as well as being located within the 65\% probability area that was surveyed. The efficiency of each model, represents the fraction of light curves that are detected to be brighter than our five-sigma limiting magnitude at the time of DECam observations.

Next, we used these simulations to examine the color magnitude space for both KN and SN (Figure~\ref{fig:colormag}). 
For this analysis, we use both KN and SN simulations. Here we require the detected object is brighter than our five-sigma limiting magnitude. Additionally, we require the object's host-galaxy photometric redshift is consistent with the LVC luminosity distance posterior at the 3$\sigma$ confidence level. Additionally, the simulated SNe were distributed in redshift according to to measured volumetric rates of SNe-Ia and SNe-CC.

\subsection{Implications for Search Efficiency}
\begin{figure*}[htp]
\centering
\includegraphics[width=1.0\linewidth]{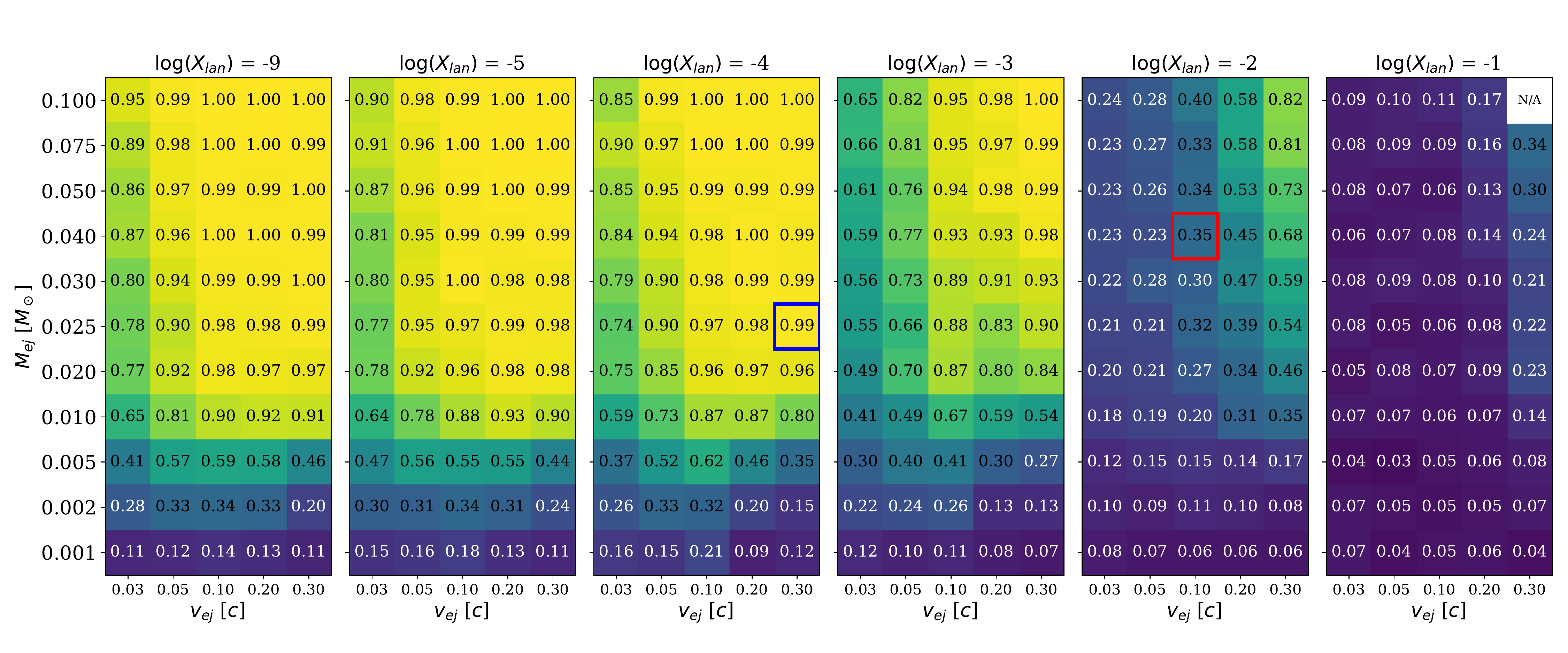}
\caption{Summary of detectable kilonovae 
given the observing conditions of May 11th, 2019. Simulations are set within the LVC distance range of S190510g. Parameters determining the components of the KN, ejecta mass ($M_{\rm ej}$), ejecta velocity ($v_{\rm ej})$, and the log of lanthanide ($\log{\rm X_{\rm lan}}$) fraction, are taken from \citep{kasen2017origin}. The coloring and labeling in each box denotes how likely we would be able to detect a KN with the given parameters assuming the event is within our observations. The box labeled ``N/A" is a combination of parameters not available in the \citealt{kasen2017origin} parameters. Additionally, we highlight the set of parameters that were identified as the likely red and blue component of GW170817 as red and blue boxes.
\label{fig:robplot}}
\end{figure*}

  Figure~\ref{fig:robplot} shows likelihood that we would have been able to detect a KN produced by this event given the observing conditions and depth of observations. Here we show all possible sets of KN parameters and note that a GW170817-like KN follows a two component model, red and blue, where the blue component is dominant at early times in the light curve evolution. Assuming S190510g is a GW170817-like KN located within our exposures, our simulations show that we would have a 99\% chance of detecting the counterpart KN. However, a wide range of KN models would have been outside of our sensitivity range and thus unobservable.

\begin{figure*}[hbt]
\centering
\includegraphics[width=0.9\linewidth]{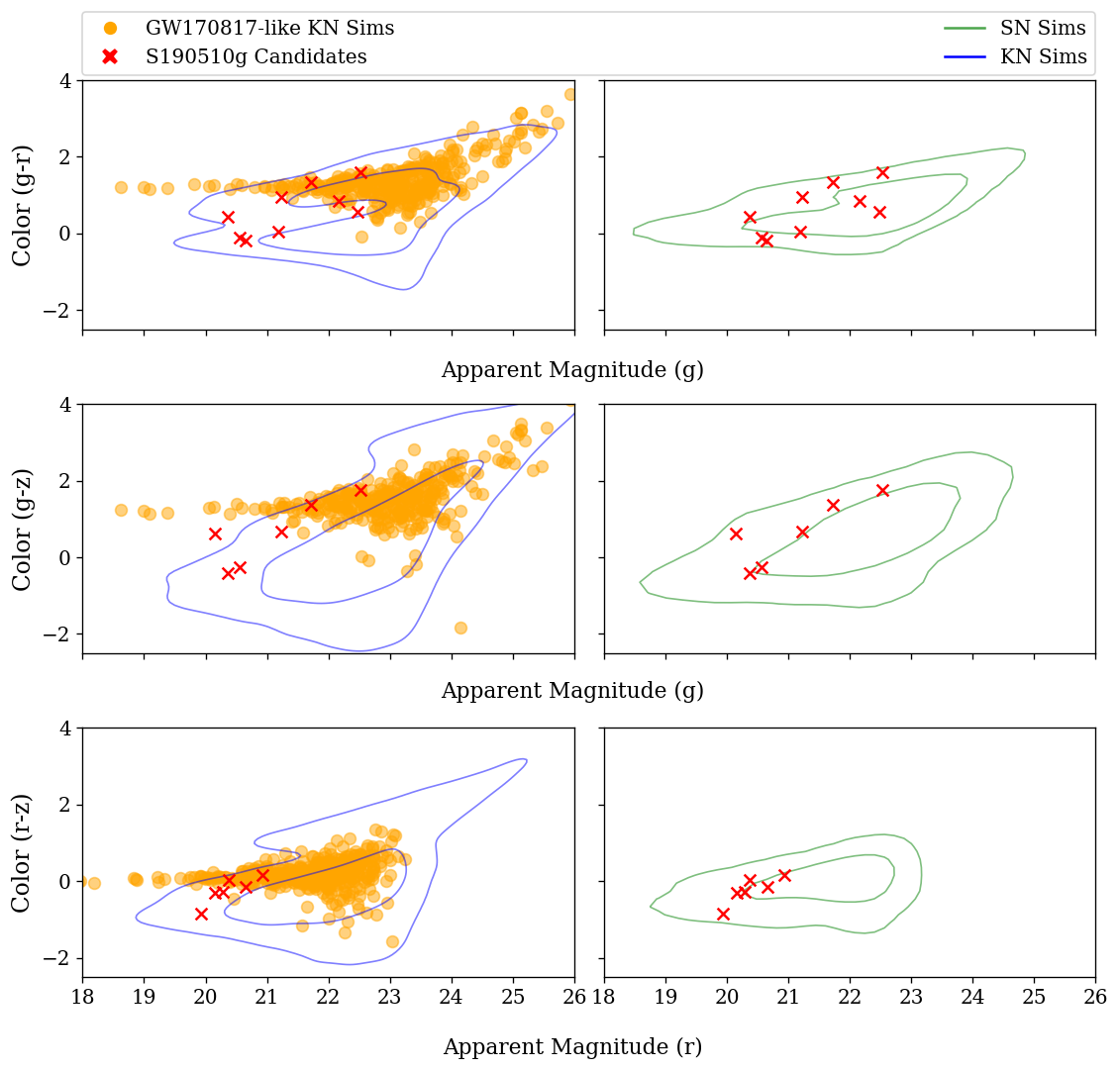}
\caption{DESGW candidates (red) as compared to where kilonovae and supernovae would be expected to live in color magnitude space given the observing conditions of our observations (i.e., 36 hours after merger, sky brightness, etc.). All simulations run using SNANA \citep{kessler2009snana}. Kilonovae simulations were generated with a burst date consistent with that reported by LVC, and at a distance consistent with the LVC distance distribution (blue contours) . The KN parameters, ejecta velocity, mass, and lanthanide fraction are randomly selected from the parameters described by Kasen 2017. Simulations with the same Kasen parameters as the blue component of GW170817 (ejecta velocity = 0.3c, lathinide fraction = $10^{-4}$ and ejecta mass =$0.025M\odot$) are shown as orange dots. Supernovae simulations consist of Type Ia and CC SN, and are generated using a peak date ranging 4 months centered around May 10th, with redshift also consistent with S190510g's distance distribtion. The contours show 50\% and 90\% density of simulations.} \label{fig:colormag}
\end{figure*}

 While we have the ability to detect such a source, it is challenging to determine a candidate to be KN or SN with a single night of observations in the absence of spectroscopic information. 
To demonstrate the difficulty of this task we examined the color magnitude space of the simulated KN and SN events. All KN simulations are shown as the blue contours in the left panel of Figure~\ref{fig:colormag}, with the parameters for the blue component of GW170817 (ejecta velocity = 0.3c, lathinide fraction = $10^{-4}$ and ejecta mass =$0.025M\odot$) highlighted as orange dots. Meanwhile the color magnitude distribution of SN simulations is shown by the green contour on the right panel of Figure~\ref{fig:colormag}. All DESGW S190510g candidates from this event (depicted as red crosses in Figure~\ref{fig:colormag}) fall within the possible $90\%$ color-magnitude regions of SN events. For a KN roughly one day after burst, and given only this color-magnitude information, each of these candidates could be either a SN or KN.

\subsection{Implications for Follow Up Strategy}
 In the first half of the O3 observing run, most of the events that included a neutron star did not have a good localization (i.e. hundreds of $\text{deg}^2$) as well as being far away ($>$200 Mpc) when compared to GW170817. While it would be ideal to cover 100\% of the localization area with multiple filters, limited telescope time and poor localization maps make this very challenging. In the following, we show that prioritizing sufficiently deep images as opposed to covering large areas and/or using multiple filters, will result in a higher chance of detecting counterparts. 
 
 To show how many events it would take to have a $50\%$ (99$\%$) chance of detecting one counterpart, we have to consider the cumulative probability inside the LVC localization map that was observed ($\Sigma_{\rm spatial}$), the fraction of DECam that was live during observations ($\epsilon_{\rm camera}$), the probability that the event is astrophysical in nature ($\epsilon_{\rm real}$), and our likelihood of being able to detect a KN at that distance given the observing conditions (i.e. the fraction of simulated lightcurves that are brighter than our five-sigma depth) ($\epsilon_{\rm efficiency}$)  
 
 \begin{equation}
     P_i = \Sigma_{\rm spatial} \times \epsilon_{ \rm camera} \times \epsilon_{\rm efficiency} \times \epsilon_{\rm real}
 \end{equation}
 \begin{equation}
     P_{\rm one} = 1 - \prod_{i}^{N}(1 - P_i)
 \end{equation}
 
 Here, $P_i$ is the probability of being able to detect a KN from a single GW event. $P_{\rm one}$ is the cumulative probability of being able to detect a single counterpart given $N$ GW events \citep{annis2016experimentally}. 
 For this calculation, we find that if we assume there is a kilonova associated with S190510g that is GW170817-like, i.e. ($\epsilon_{\rm efficiency}$ = 0.993), we would need to observe 3 (19) identical events with $\Sigma_{\rm spatial}$ = 0.65, $\epsilon_{\rm camera}$ = 0.8, $\epsilon_{\rm real}$ = 0.42, and $\epsilon_{\rm efficiency}$ = 0.993 in order to have $50\%$ (99\%) probability of identifying the event using the current strategy. Since there is no way of knowing that the event will have a lightcurve similar to GW170817, we also calculate this using the average efficiency value of all KN parameters, $\epsilon_{\rm efficiency} = 0.553 $ with all other parameters the same. Here we find that we would need 6 (36) events to reach $50\%$ ($99\%$) likelihood of detecting the counterpart. 
 
 We then repeat this calculation assuming the observing strategy uses one filter instead of three. If we conserve the telescope time used and area surveyed per event, we can then increase the exposure time from 40 seconds to 170 seconds. In this scenario, the efficiency for a GW170817-like KN is 0.995, meaning we would again need 3 (19) events to have $50\%$ ($99\%$) likelihood of detection. Using the average efficiency in this scenario though, 0.742, we would only need 4 (27) events to have $50\%$ ($99\%$) likelihood of detecting a counterpart. By increasing the depth of our observations, we become sensitive to more KN models and will thus need to observe fewer total GW events to have a high probability of making a detection.

\section{Conclusion}

We performed a follow up analysis of the GW trigger S190510g, using DECam target of opportunity time data from May 11th 2019. 
We demonstrated the DESGW team's ability to quickly process new images in real time, averaging $\sim1$hr for image processing to complete. The final DESGW candidate list is summarized in Table 1, with five candidates, desgw-190510a, c, i, j, and k being ruled out due to secondary follow up efforts by KMTNet, Swift-XRT, and Magellan. Similarly, candidates desgw-190510b and h have been identified based on previous observation as recorded in the Transient Name Server. This leaves 4 candidates from the DESGW candidate list that were not classified by secondary follow-up. Each of these candidates have color information that is consistent with SN.

Additionally, we used simulated KN to show the efficiency of detecting a KN counterpart given the observing conditions of the observations to find that we have a 99\% chance of being able to detect a KN counterpart assuming the light curve has the same physical parameters as GW170817 using the \cite{kasen2017origin} model (Fig.~\ref{fig:robplot}) within our observations. However, this efficiency is not uniform across all KN models. We also used KN and SN simulations to study where in color magnitude space they land. We find that all of our candidates are consistent with both KN and SN using this metric.  

To make ourselves more sensitive to all KN models, we suggest prioritizing longer exposure times over multiple filters and covering large portions of the localization area for future observations. Using exposures that are ~4 times longer than those used for this follow up, we would only need to observe 4 events (identical to S190510g) to have a 50\% chance of detecting a KN counterpart within the $65\%$ probability region observed and with these observing conditions, compared to the 6 events needed using the current strategy.

\section*{Acknowledgments}
Funding for the DES Projects has been provided by the U.S. Department of Energy, the U.S. National Science Foundation, the Ministry of Science and Education of Spain, the Science and Technology Facilities Council of the United Kingdom, the Higher Education Funding Council for England, the National Center for Supercomputing 
Applications at the University of Illinois at Urbana-Champaign, the Kavli Institute of Cosmological Physics at the University of Chicago, 
the Center for Cosmology and Astro-Particle Physics at the Ohio State University,
the Mitchell Institute for Fundamental Physics and Astronomy at Texas A\&M University, Financiadora de Estudos e Projetos, 
Funda{\c c}{\~a}o Carlos Chagas Filho de Amparo {\`a} Pesquisa do Estado do Rio de Janeiro, Conselho Nacional de Desenvolvimento Cient{\'i}fico e Tecnol{\'o}gico and 
the Minist{\'e}rio da Ci{\^e}ncia, Tecnologia e Inova{\c c}{\~a}o, the Deutsche Forschungsgemeinschaft and the Collaborating Institutions in the Dark Energy Survey. 

The Collaborating Institutions are Argonne National Laboratory, the University of California at Santa Cruz, the University of Cambridge, Centro de Investigaciones Energ{\'e}ticas, 
Medioambientales y Tecnol{\'o}gicas-Madrid, the University of Chicago, University College London, the DES-Brazil Consortium, the University of Edinburgh, 
the Eidgen{\"o}ssische Technische Hochschule (ETH) Z{\"u}rich, 
Fermi National Accelerator Laboratory, the University of Illinois at Urbana-Champaign, the Institut de Ci{\`e}ncies de l'Espai (IEEC/CSIC), 
the Institut de F{\'i}sica d'Altes Energies, Lawrence Berkeley National Laboratory, the Ludwig-Maximilians Universit{\"a}t M{\"u}nchen and the associated Excellence Cluster Universe, 
the University of Michigan, NFS's NOIRLab, the University of Nottingham, The Ohio State University, the University of Pennsylvania, the University of Portsmouth, 
SLAC National Accelerator Laboratory, Stanford University, the University of Sussex, Texas A\&M University, and the OzDES Membership Consortium.

Based in part on observations at Cerro Tololo Inter-American Observatory at NSF’s NOIRLab (NOIRLab Prop. ID 2012B-0001; PI: J. Frieman), which is managed by the Association of Universities for Research in Astronomy (AURA) under a cooperative agreement with the National Science Foundation.

The DES data management system is supported by the National Science Foundation under Grant Numbers AST-1138766 and AST-1536171.
The DES participants from Spanish institutions are partially supported by MICINN under grants ESP2017-89838, PGC2018-094773, PGC2018-102021, SEV-2016-0588, SEV-2016-0597, and MDM-2015-0509, some of which include ERDF funds from the European Union. IFAE is partially funded by the CERCA program of the Generalitat de Catalunya.
Research leading to these results has received funding from the European Research
Council under the European Union's Seventh Framework Program (FP7/2007-2013) including ERC grant agreements 240672, 291329, and 306478.
We  acknowledge support from the Brazilian Instituto Nacional de Ci\^encia
e Tecnologia (INCT) e-Universe (CNPq grant 465376/2014-2).

This manuscript has been authored by Fermi Research Alliance, LLC under Contract No. DE-AC02-07CH11359 with the U.S. Department of Energy, Office of Science, Office of High Energy Physics.

This material is based upon work supported by the National Science Foundation Graduate Research Fellowship Program under Grant No. 1744555. Any opinions, findings, and conclusions or recommendations expressed in this material are those of the author(s) and do not necessarily reflect the views of the National Science Foundation.

R. Morgan thanks the LSSTC Data Science Fellowship Program, which is funded by LSSTC, NSF Cybertraining Grant \#1829740, the Brinson Foundation, and the Moore Foundation; his participation in the program has benefited this work.

F.O.E.\ acknowledges support from the FONDECYT grant nr.\ 1201223.

\bibliography{AGbib}
\bibliographystyle{yahapj}
\end{document}